\documentclass{PoS}
\usepackage{graphicx}
\usepackage{verbatim}
\usepackage{amssymb}
\usepackage{dsfont}
\usepackage{amsmath}
\usepackage{eufrak}
\usepackage[T1]{fontenc}

\title{Non-associative star products and quantization of non-geometric backgrounds in string and M-theory}

\ShortTitle{Non-associative star products}

\author{\speaker{Vladislav G. Kupriyanov}\\
         {\it Max-Planck-Institut f\"ur Physik,
  Werner-Heisenberg-Institut, M\"unchen, Germany
}
\\ {\it CMCC-Universidade Federal do ABC, Santo Andr\'e, SP, 
Brazil}\\  {\it 
Tomsk State University, Tomsk, Russia}
\\

        E-mail: {vladislav.kupriyanov@gmail.com}}

\abstract{We review two known in the literature exemples of non-associative star products. The first one is the phase space star product representing quantization of non-geometric $R$-flux background in closed string theory. The second is the octonionic star product which provides the quantization of the quasi-Poisson algebra isomorphic to the Malcev algebra of imaginary octonions. We discuss in details the construction, properties and physical applications of these star products. In particular, we consider the quantization of non-geometric M-theory background. Based on these two exemples we formulate the minimal set of physically motivated conditions for the definition of non-associative deformation quantization. }

\FullConference{Corfu Summer Institute 2017 'School and Workshops on Elementary Particle Physics and Gravity'\\
		2-28 September 2017\\
		Corfu, Greece}

\begin{document}

\section{Introduction}

The standard Hamiltonian description of the classical mechanical system is based on the notion of the {\it Poisson} bracket of two functions $\{f,g\}$, which is bi-linear, antisymmetric and satisfies the Jacobi identity:
\begin{equation}
\{f,g,h\}:=\{f,\{g,h\}\}+\{h,\{f,g\}\}+\{g,\{h,f\}=0\ .\label{jac}
\end{equation}
An important advantage of this description is a Poisson Theorem, stating that the Poisson bracket of two integrals of motion is again an integral of motion (not necessarily new), i.e., in terms of the Poisson bracket one may formulate the algebra of classical observables. The other essential point is that the standard Hamiltonian formalism is compatible with the canonical quantization scheme.

However, there are several physical situations when the system is formulated in terms of bracket violating the Jacobi identity (\ref{jac}). To start with, let us remind that the magnetic field $\vec B$ can be introduced in the theory via the algebra of coordinates and covariant momenta: 
\begin{equation}
\{x^i, x^j\}=0,\qquad \{x^i,\pi_j\}=\delta^i_j,\qquad \{\pi_i,\pi_j\}=e\varepsilon_{ijk}B^k( x).\label{monopole}
\end{equation}
The Jacobiator of the three covariant momenta reads
\begin{equation}
\{\pi_i,\pi_j,\pi_k\}=e\,\varepsilon_{ijk}\,\mbox{div}\,\vec B\,.
\end{equation}
If one admits the presence of the magnetic charges of the density $\rho(x)$, then $\mbox{div}\,\vec B=\rho(x)\neq0$, meaning the violation of the Jacobi identity. Sometimes in the literature the algebra (\ref{monopole}) is called as a monopole algebra. Throughout this paper we will also call the antisymmetric bi-linear bracket $\{f,g\}$, for which the Jacobi identity does not hold as a {\it quasi-Poisson} structure.

Another exemple quasi-Poisson brackets comes from the closed string theory with non-geometric fluxes. In particular, the constant $R$-flux algebra \cite{BP,Lust:2010iy,Blumenhagen:2011ph} reads:
\begin{eqnarray}
\{x^i, x^j\}=\mbox{$\frac{\ell_s^3}{\hbar^2}$} \, R^{ijk} \, p_k \ ,  \qquad \{x^i,p_j\}=\delta^i_j \qquad  \{p_i,p_j\}=0 \ ,\label{flux}
\end{eqnarray}
with $R^{ijk}=R\, \varepsilon^{ijk}$, and $\varepsilon^{123}=+1$. Note that making, $p\to x$ and $x\to-p$, one obtains the monopole algebra corresponding to a constant magnetic charge distribution $\rho(x)=R$. There are also interesting exemples of non-Jacobian structure related to the non-geometric backgrounds in M-theory \cite{GLM}, which will be discussed in the details in the Section 4, as well as its magnetic analogous \cite{LMS}. In the context of open string theory it was pointed out in \cite{CoSch,MK1,Herbst}, that the presence of a non-constant $B$-field, may result in the non-associativity of a star product. For explicit exemples of non-commutative and non-associative structures see \cite{BBKL} and references therein.

The problem we are going to discuss in this contribution to the Proceedings of the Corfu Summer Institute is the consistent quantization of quasi-Poisson structures in the framework of the deformation quantization.
More precisely, we are interested in the quantization of the {\it quasi-Poisson} bracket
\begin{equation}
\{f,g\}=\theta^{ij}(x)\ \partial_i f\ \partial_jg\, \label{PB}
\end{equation}
without imposing the Jacobi identity on the bi-vector $\theta^{ij}$, i.e., in general,
\begin{equation}
\Pi^{ijk}:=\frac{1}{3}\left(\theta^{il}\partial_l\theta^{jk}+ \theta^{kl}\partial_l\theta^{ij}+\theta^{jl}\partial_l\theta^{ki}\right)\neq0\ .\label{jacobi}
\end{equation}
In the standard approach to the deformation quantization \cite{BFFLS} the star product is defined as a formal expansion,
\begin{equation}
 f\star g = f \cdot g +\sum_{r=1}^\infty (i\hbar)^r C_r(f,g) \ ,\label{star}
\end{equation}
where $\hbar$ is a deformation parameter, and $C_r(f,g)$ denote some bi-differential operators. The ``initial condition'', consisting in the requirement that in the quasi-classical limit the star commutator should reproduce the given quasi-Poisson bracket,
\begin{equation}
 \lim_{\hbar\to0}\, \frac{[f,g]_{\star}}{2i\hbar}=\{f,g\}\,,\label{ic}
\end{equation}
guarantees that the star product (\ref{star}) provides the quantization of the classical bracket (\ref{PB}).

Remind that in the standard setting the problem of deformation quantization is the problem of the construction of the star product $f\star g$, which should satisfy the condition of associativity,
\begin{equation}
A(f,g,h):=(f\star g)\star h-f\star(g\star h)=0\ ,\label{ass}
\end{equation}
and represent the deformation of the ordinary point-wise product, $f\cdot g$, in the direction of a given bi-vector field $\theta^{ij}(x)$, i.e., it should satisfy (\ref{star}) and (\ref{ic}). There are natural "gauge" transformations acting on star products and defined as: $\mathcal{D}:\star\,\to\,\star'$,
\begin{equation}
f\star' g= \mathcal{D}^{-1}\big( \mathcal{D}f\star \mathcal{D}g \big) \ ,\label{gauge}
\end{equation}
with $\mathcal{D}=1+\mathcal{O}\left( \hbar \right)$, being an invertible differential operator. If $\star$ is associative and represents the quantization of the bracket $\{f,g\}$, then $\star'$ is also associative and gives quantization of the same bracket. Physically gauge equivalent star products represent different quantizations of the same classical bracket $\{f,g\}$.

In the zero order in deformation parameter the condition (\ref{ass}) is trivial. In the first order one obtains the equation on $C_1(f,g)$,
\begin{equation}
C_1(f,g)\ h+C_1(fg,h)-C_1(f,gh)-f\ C_1(g,h)=0\ .
\end{equation}
The symmetric part of the bi-differential operator $C_1$, $C_1^+(f,g)=C_1^+(g,f)$, can be killed by a gauge transformation (\ref{gauge}), while the antisymmetric one, $C_1^-(f,g)=-C_1^-(g,f)$, is defined from the initial condition (\ref{ic}), and is given by $C_1^-(f,g)=\{f,g\}/2$. The associativity equation (\ref{ass}) in the second order $\mathcal{O}\left( \hbar^2 \right)$, implies the consistency condition requiring that the classical Jacobiator of any three functions defined in (\ref{jac}) should vanish.
It happens that no other non-trivial conditions on the classical bracket $\{f,g\}$ appear. The existence of the associative star product for any Poisson bi-vector $\theta^{ij}$ follows from the Formality Theorem by Kontsevich \cite{Kontsevich}, which also states that each two star products $\star$ and $\star'$ which quantize the same Poisson bracket $\{f,g\}$ are related by the gauge transformation (\ref{gauge}).

If now we relax the Jacoby identity, i.e., suppose that $\Pi^{ijk}\neq0$, the associativity equation (\ref{ass}) will be violated already in the second order in $\hbar$, and the question is which condition should be used instead of the associativity (\ref{ass}) to restrict the higher order terms in (\ref{star}). Or, in other words, how to keep the violation of the associativity under control?

The previous attempt to answer this question consisted in the requirement of the \emph{alternativity} of the star product \cite{KupPos}. Remind that the multiplication is called alternative if the corresponding associator (\ref{ass}) is alternating, i.e., antisymmetric in all arguments. In this case the associator is proportional to the jacobiator. The alternativity of the star product at least on the some class of functions is useful, e.g., for the definition of states in non-associative quantum mechanics \cite{Bojowald2014,Bojowald2015}. However, later on it was shown that the monopole star products are not alternative \cite{Bojow}. In \cite{Kup24} it was demonstrated that for alternative deformation quantization the quasi-classical limit of the star commutator has to satisfy the Malcev identity.  In \cite{KS} we proved that neither (\ref{flux}), nor the quasi-Poisson structure isomorphic to the algebra of the imaginary octonions fail to be Malcev-Poisson structure and consequently do not allow the alternative star products. In \cite{Vassilevich:2018gkl} it was shown more stronger statement: the quasi-Poisson structure satisfies the Malcev identity only if it is Poisson obeying (\ref{jac}). From the practical point of view it means that non-trivial (non-associative) alternative star products do not exist.

In this notes we consider two known in the literature examples of the non-associative star products: the phase space star product for the constant $R$-flux \cite{MSS2} and the octonionic star product \cite{Kup24}. Discussing the properties of these star products and some physical applications, like the formulation of the non-associative quantum mechanics \cite{Mylonas2013} the quantization of non-geometric M-theory backgrounds \cite{KS} we will formulate the physically motivated minimal set of requirements for the non-associative deformation quantization. Finally we briefly discuss how one may proceed in constructing a star product satisfying these conditions for the arbitrary given quasi-Poisson structure $\theta^{ij}(x)$.

\section{Constant $R$-flux in closed string theory and quantization}

Originally the star product providing the quantization of the quasi-Poisson structure (\ref{flux}) first appeared in
\cite{MSS2} where it was derived using the Kontsevich formula for
deformation quantization of twisted Poisson structures. In these notes to construct the star product we follow an approach based on associative algebra of differential operators \cite{KV15}. The significance and utility of this star product in understanding non-geometric string theory and non-associative quantum mechanics is exemplified in~\cite{MSS2,BaLu,Mylonas2013,Aschieri2015}.

For the convenience  let us first represent the constant $R$-flux algebra (\ref{flux}) in the form
\begin{eqnarray}
\{x^I, x^J\}=\Theta^{IJ}(x)=\begin{pmatrix}
  \frac{\ell_s^3}{\hbar^2}\, R^{ijk}\, p_k &  -\delta^i_j   \\
  \delta^i_j  & 0 \end{pmatrix}
\qquad \mbox{with}\qquad  x=(x^I) =({\bf x},{\bf p}).\label{rb}
\end{eqnarray}
A star product is defined by associating a (formal) differential operator $\hat f$ to a function $f$ as
\begin{equation}
(f\star g)(x)=\hat f \triangleright g(x)~,
\label{d3}
\end{equation}%
where the symbol $\triangleright$ denotes the action of a differential operator on
a function. In particular one has
\begin{equation}\label{op1}
x^I\star f=\hat x^I\triangleright f(x)~,
\end{equation}%
where the operators
\begin{equation}\label{op2}
\hat x^I=x^I+\mbox{$\frac{i\hbar}2$}\, \Theta^{IJ}(x)\, \partial_J\ ,
\end{equation}
with $\partial_i=\frac{\partial}{\partial x^i}$ and $\partial_{i+3}=\frac{\partial}{\partial p_i}$ for $i=1,2,3$. One may easily check that
\begin{equation}
[\hat x^i,\hat x^j]=\mbox{$\frac{i\ell_s^3}\hbar$}\, R^{ijk}\,\label{eq:stringhatxcomm}
\big(\hat p_k+i\hbar\, \partial_k \big) \ ,
\end{equation}
which means that the operators (\ref{op2}) do not close an algebra.
Taking (\ref{op1}) and (\ref{op2}) as a definition of the star product, one may easily calculate for the star commutator and the star jacobiator
\begin{equation}
[x^I,x^J]_\star:= x^I\star x^J-x^J\star x^I=i\hbar\, \Theta^{IJ}
\qquad \mbox{and} \qquad [x^i,x^j,x^k]_\star =-3\, \ell_s^3\,R^{ijk} \ ,
\label{eq:Rbrackets}
\end{equation}
which thereby provide a quantization of the classical brackets \eqref{rb}.

To define the star product $f\star g$ between two arbitrary functions on phase space, we introduce the notion of \emph{Weyl star product} by requiring that, for any $f$, the differential operator $\hat f$ defined by (\ref{d3})
can be obtained by symmetric ordering of the operators $\hat x^I$. Let
$\tilde f(k)$ denote the Fourier transform of $f(x)$, with $k=(k_I) =({\bf  k},{\bf  l})$ and ${\bf  k}=(k_i),{\bf  l}=(l^i) \in \mathbb{R}^3$. Then
\begin{equation}
\hat f=W( f)  :=\int\, \frac{d^{6}k}{( 2\pi ) ^{6}} \ 
\tilde{f}(k) \, e^{-i k_{I}\, \hat{x}^{I}} \ .  \label{2}
\end{equation}
For example, $W(x^I\, x^J)=\tfrac 12\, (\hat x^I\,\hat x^J + \hat x^J\, \hat x^I)$. 
Weyl star products satisfy
   \begin{equation}\label{weyl}
    \big(x^{I_1}\cdots x^{I_n}\big)\star f=\frac 1{n!} \, \sum_{\sigma\in S_n}\, x^{I_{\sigma(1)}}\star\big(x^{I_{\sigma(2)}}\star \cdots \star (x^{I_{\sigma(n)}}\star f)\cdots\big)\ ,
\end{equation}
where the sum runs over all permutations in the symmetric group $S_n$ of degree $n$. It should be stressed that because of the (\ref{eq:stringhatxcomm}), in general, $\widehat{ f\star g} \ne \hat f \circ \hat g$, meaning also that the star product should not be associative.

To obtain an explicit form for the corresponding star product we first observe that since $[k_{I}\, {x}^{I}, k_{J}\, \Theta^{JL}\, \partial_L]=0$ one can write
\begin{equation*}
e^{-i k_{I}\, \hat{x}^{I}} =e^{- ik\,\cdot\, {x}}\,
e^{\frac\hbar2\, k_{I}\, \Theta^{IJ}(x) \, \partial_J} \ ,
\end{equation*}
with $\,\cdot\,$ denoting the standard Euclidean inner product of vectors. By the relation $k_{I}\, k_{L}\, \Theta^{IJ}\, \partial_J
\Theta^{LM}\, \partial_M=\frac{ell_s^3}{\hbar^2}\, k_i\, k_l\, R^{lki}\, \partial_k=0$ it follows that
\begin{equation*}
\big(k_{I}\,\Theta^{IJ}\,\partial_J\big)^n=k_{I_1}\cdots k_{I_n}\, \Theta^{I_1J_1}\cdots\Theta^{I_nJ_n}\, \partial_{J_1}\cdots\partial_{J_n} \ .
\end{equation*}
One may also write
\begin{equation*}
\big(\overleftarrow{\partial}_{I}\, \Theta^{IJ}\, \overrightarrow{\partial}_J\big)^n=\overleftarrow{\partial}_{I_1}\cdots \overleftarrow{\partial}_{I_n}\, \Theta^{I_1J_1}\cdots\Theta^{I_nJ_n}\, \overrightarrow{\partial}_{J_1}\cdots\overrightarrow{\partial}_{J_n} \ ,
\end{equation*}
where $\overleftarrow{\partial}_I$ and $\overrightarrow{\partial}_I$
stand for the action of the derivative $\frac\partial{\partial x^I}$
on the left and on the right correspondingly. Thus the Weyl star
product representing quantization of the quasi-Poisson bracket
(\ref{rb}) can be written in terms of a bidifferential operator as
\begin{equation}
(f\star g)(x) =\int\, \frac{d^{6}k}{( 2\pi) ^{6}} \ 
\tilde{f}(k) \, e^{-i k_{I}\, \hat{x}^{I}}\triangleright
g(x)=f(x)\, e^{\frac{i\hbar}{2}\, \overleftarrow{\partial}_{I}\,
  \Theta^{IJ}(x) \, \overrightarrow{\partial}_J} \, g(x) \ . \label{starrb}
\end{equation}
It is easy to see that (\ref{starrb}) is Hermitean, $(f\star
g)^\ast=g^\ast\star f^\ast$, and unital, $f\star1=f=1\star f$. Moreover it is 2-cyclic or closed,
\begin{equation}
\int \,   \ f\star g = \int \,  \ f\cdot g \ ,\label{cust1}
\end{equation}
and 3-cyclic, meaning that the integrated associator vanishes,
\begin{equation}
\int\,   (f\star g)\star h = \int \, 
f\star(g\star h) \ ,\label{aust1}
\end{equation}
see~\cite{Mylonas2013} for the proof.

For later use, let us rewrite the star product $f\star g$ in integral
form through the Fourier transforms $\tilde{f}$ and $\tilde{g}$
alone,
\begin{equation}\label{starrb1}
(f\star g)(x) =\int\, \frac{d^{6}k}{( 2\pi
) ^{6}}\ \frac{d^{6}k'}{( 2\pi
) ^{6}}\ \tilde{f}(k)\, \tilde{g}( k'\, )\, e^{i\mathcal{ B}(k,k'\,
)\,\cdot\, x} \ ,
\end{equation}
where
\begin{equation}\label{BR}
\mathcal{ B}( k, k'\,)\,\cdot\, x :=( {\bf k}+ {\bf k'}\, )\,\cdot\,  {\bf x}+ (
{\bf l}+ {\bf l'}\,)\,\cdot\,  {\bf p}-\mbox{$\frac{\ell_s^3}{2\hbar}$}\, R\, {
  {\bf p}}\,\cdot\,({\bf  k}\,\times_\varepsilon \,{\bf 
  k}'\, )+\mbox{$\frac{\hbar}{2}$}\,\big({\bf  l}\,\cdot\, {\bf  k}'
-{\bf  k}\,\cdot\, {\bf  l}'\, \big) \ ,
\end{equation}
with
$( {\bf k}\,\times_\varepsilon \, {\bf k'}\,)_i=\varepsilon_{ijl}\, k_j\, k'_l$ the usual cross product of three-dimensional vectors.

\section{Octonions and M-theory $R$-flux background}

The algebra $\mathbb{O}$ of octonions is the best known example of a nonassociative but alternative algebra. Every octonion $X\in \mathbb{O}$ can be written in the form
\begin{equation}\label{oct}
X=k^0\,{\bf 1}+ k^A\,e_A
\end{equation}
where $k^0,k^A\in\mathbb{R}$, $A=1,\dots,7$, while ${\bf 1}$ is the identity element and the imaginary unit octonions $e_A$ satisfy the multiplication law
\begin{equation}\label{oct1}
e_A\, e_B=-\delta_{AB}\,{\bf 1} +\eta_{ABC}\, e_C \ .
\end{equation}
Here $\eta_{ABC}$ is a completely antisymmetric tensor of rank three
with non-vanishing values
\begin{equation}
\eta_{ABC}=+1 \qquad \mbox{for} \quad ABC = 123 , \ 435, \ 471, \ 516,
\ 572, \ 624, \ 673 \ .
\end{equation}
The algebra $\mathbb{O}$ is neither commutative nor associative. The commutator algebra of the octonions is given by
\begin{equation}\label{oct2}
[e_A,e_B]:=e_A\, e_B-e_B\, e_A=2\, \eta_{ABC}\, e_C\ .
\end{equation}
The structure constants $\eta_{ABC}$ satisfy the contraction identity
\begin{equation}\label{epsilon7}
\eta_{ABC}\, \eta_{DEC}=\delta_{AD}\, \delta_{BE}-\delta_{AE}\,
\delta_{BD}+\eta_{ABDE} \ ,
\end{equation}
where $\eta_{ABCD}$ is a completely antisymmetric tensor of rank four
with nonvanishing values
$$
\eta_{ABCD}= +1 \qquad \mbox{for} \quad ABCD = 1267, \ 1346, \ 1425, \
1537, \ 3247, \ 3256, \ 4567 \ .
$$
Taking into account (\ref{epsilon7}), for the Jacobiator we get
\begin{equation}\label{oct3}
[e_A,e_B,e_C]:=[e_A,[e_B,e_C]]+[e_C,[e_A,e_B]]+[e_B,[e_C,e_A]]=-12\,
\eta_{ABCD}\, e_D \ ,
\end{equation}
and the alternative property of the algebra $\mathbb{O}$ implies that the
Jacobiator is proportional to the associator, i.e.,
$[X,Y,Z]=6\,\big((X\,Y)\, Z-X\,(Y\, Z) \big)$ for any three octonions
$X,Y,Z\in\mathbb{O}$.

Introducing $f_i:=e_{i+3}$ for $i=1,2,3$, the algebra of the imaginary octonions (\ref{oct2}) can be rewritten in components as 
\begin{eqnarray}\label{oct2a}
[e_i,e_j]&=&2\, \varepsilon_{ijk}\, e_k \qquad \mbox{and} \qquad [e_7,e_i] \ = \ 2\,
             f_i\ ,\\[4pt]
[f_i,f_j]&=&-2\, \varepsilon_{ijk}\, e_k \qquad \mbox{and} \qquad [e_7,f_i] \ = \
             -2\, e_i\ ,\nonumber\\[4pt]
[e_i,f_j]&=&2\, (\delta_{ij}\, e_7- \varepsilon_{ijk}\, f_k) \ .\nonumber
\end{eqnarray}
Defining the coordinates and momenta in terms of the imaginary octonions as
\begin{equation}
x^i= \mbox{$\frac{\sqrt{\lambda\,
      \ell_s^{3}\, R}}{2\hbar}$}\,f_i\ ,\,\,\,\,\, p_i=-\mbox{$\frac{\lambda}{2}$}\,e_i\ ,\,\,\,\,\,\,x^4=\mbox{$\frac{\sqrt{\lambda^3\, \ell_s^{3}\, R}}{2\hbar}$}\,e_7\ ,
\end{equation}
we obtain the quasi-Poisson algebra
\begin{eqnarray}\label{RM}
\{x^i,x^j\}_\lambda &=&\mbox{$\frac{\ell_s^3}{\hbar^2}$}\,
                        R^{4,ijk4}\, p_k \qquad \mbox{and} \qquad
                        \{x^4,x^i\}_\lambda \ = \ \mbox{$\frac{\lambda\, \ell_s^3}{\hbar^2}$}\, R^{4,1234}\, p^i , \\[4pt]
\{x^i,p_j\}_\lambda &=&\delta^i_j\,x^4+\lambda\,
                        \varepsilon^i{}_{jk}\, x^k \qquad \mbox{and}
                        \qquad \{x^4,p_i\}_\lambda \ = \ \lambda^2\,x_i \ , \nonumber\\[4pt]
\{p_i,p_j\}_\lambda &=&-\lambda\, \varepsilon_{ijk}\, p^k\ . \nonumber
\end{eqnarray}
Now taking the contraction limit $\lambda\to0$ we observe that the element $x^4$ becomes a central element and can be taken to be identity operator, while the algebra generated by $x^i$ and $p_i$ coincide with the constant $R$-flux algebra (\ref{flux}). The main conjecture of \cite{GLM} is that the quasi-Poisson brackets (\ref{RM}) provide the uplift of the string $R$-flux algebra to M-theory. In this sense $\lambda$ plays the role of the M-theory radius.

\section{Quantization of M-theory $R$-flux background}

In this section we discuss the problem how to construct the star product which in the quasi-classical limit would reproduce the quasi-Poisson structure (\ref{RM}). For simplicity first we quantize the algebra of the quasi-Poisson brackets
\begin{equation}
\{\xi_A,\xi_B\}_\eta=2\eta_{ABC}\xi_C\,,\qquad \mbox{with} \qquad \xi_A\in\mathbb{R}\,,\label{oct2a}
\end{equation}
isomorphic to the algebra of imaginary octonions (\ref{oct2}). The classical Jacobiator reads:
\begin{equation}\label{mp2}
\{\xi_A,\xi_B,\xi_C\}_\eta=-12\,
\eta_{ABCD}\, \xi_D \ .
\end{equation}
The key mathematical tool used for the construction of the \emph{octonionic star product} providing the quantizing the algebra (\ref{oct2a}) is the \emph{deformed vector sum}, whose properties are essentially based on the notion of the seven-dimensional cross product related to the algebra of octonions $\mathbb{O}$, see Section 4.1 for details.

To obtain the desired star product we then make the change of the coordinates,
\begin{equation}\label{oct8}
\vec x = \big(x^A\big) = \big({\bf x},x^4,{\bf p}\big) := {\mit\Lambda}\,
\vec\xi= \mbox{$\frac1{2\hbar}$}\, \big(\sqrt{\lambda\,
      \ell_s^{3}\, R}\ {\bf\sigma}\,,\,
\sqrt{\lambda^3\, \ell_s^{3}\, R}\ \sigma^4\,,\, -\lambda\,\hbar\ {\bf\xi} \big) \ ,
\end{equation}
where the $7\times7$ matrix ${\mit\Lambda}$ is given by
\begin{equation}\label{oct6}
{\mit\Lambda} = \big({\mit\Lambda}^{AB}\big)= \frac1{2\hbar} \,
\begin{pmatrix}
0 & {\sqrt{\lambda\, \ell_s^3\, R}}\ {\bf 1}_3 & 0 \\
0 & 0 & \sqrt{\lambda^3\, \ell_s^3\, R} \\
 -\lambda\, \hbar\ {\bf 1}_3 & 0 & 0
\end{pmatrix}
\end{equation}
with ${\bf 1}_3$ the $3\times3$ identity matrix, and $\sigma^i:=\xi_{i+3}$ for
$i=1,2,3$ and $\sigma^4:= \xi_7$. From the classical brackets (\ref{oct2a}) one obtains the quasi-Poisson algebra
\begin{equation}\label{oct4a}
\{x^A,x^B\}_\lambda=2\,\lambda^{ABC} \, x^C \qquad \mbox{with} \quad \lambda^{ABC}:= {\mit\Lambda}^{AA^\prime}\,{\mit\Lambda}^{BB^\prime}\,\eta_{A^\prime B^\prime C^\prime}\, {\mit\Lambda}^{-1}_{C^\prime C} \ ,
\end{equation}
which can be written in components as (\ref{RM}).

The consistency check is the calculation of the contraction limit $\lambda\to0$ in the obtained star product giving the quantization of (\ref{oct4a}), which according to the logic of the previous Section should reproduce the expression (\ref{starrb1}) for the star product of the closed string $R$-flux algebra.

\subsection{Deformed vector sums}

To warm up let us first discuss the following problem. How to define a binary operation $\circledast_\varepsilon$ on the unit ball $|{\bf q}\,|\leq1$ in $\mathbb{R}^3$, satisfying two main properties: it should be associative and its commutator should reproduce the cross product in three dimensions. 

The solution was given in \cite{Freidel2005}. For each ${\bf q},{\bf q}\,^{\prime}$ from the
unit ball in $\mathbb{R}^3$, define the map called \emph{vector star sum} by the rule:
\begin{equation}\label{ds3}
{\bf q}\circledast_\varepsilon{\bf q}\,^{\prime}=  \epsilon_{{\bf q},{\bf q}\,'}\,\big(\, \sqrt{1-| {\bf q}\,^{\prime}|^2}\,\,{\bf q}+ {\sqrt{1-| {\bf q}\,|^2} }\,\,{\bf q}\,^{\prime}-{\bf q}\,\times_\varepsilon\, {\bf q}\,^{\prime}\big) \ ,
\end{equation}
with $\epsilon_{{\bf q},{\bf q}\,'}=\pm1$ being the sign of $ \sqrt{1-| {\bf q}\,^{\prime}|^2}\,\,{\bf q}+ {\sqrt{1-| {\bf q}\,|^2} }\,\,{\bf q}\,^{\prime}-{\bf q}\,\times_\varepsilon\, {\bf q}\,^{\prime}$. Using the properties of three-dimensional cross product one may easily check that the operation $\circledast_\varepsilon$ enjoys the following properties.
\begin{description}
\item[(V1) ] Vector ${\bf q}\circledast_\varepsilon{\bf q}\,^{\prime}$ belongs to the unit ball in $\mathbb{R}^3$,
\begin{equation}
1-|{\bf q}\circledast_\varepsilon{\bf q}\,^{\prime}\,|^2=\big(\, \sqrt{1-|{\bf q}\, |^2}\, \sqrt{1-|{\bf q}\,^{\prime}\, |^2}-{\bf q}\,\cdot\,{\bf q}\,^{\prime}\, \big)^2\geq0;
\end{equation}
\item[(V2) ] Commutator reproduces the cross product, 
\begin{equation}
{\bf q}\circledast_\varepsilon{\bf q}\,^{\prime}-{\bf q}\,^{\prime}\circledast_\varepsilon{\bf q}=\frac{1}{2} \,{\bf q}\,^{\prime}\times_\varepsilon{\bf q};
\end{equation}
\item[(V3) ] It is associative,
\begin{equation}\label{vssass}
 {\bf A}({\bf q},{\bf q}\,^{\prime},{\bf q}\,^{\prime\prime}\,):= ({\bf q}\circledast_\varepsilon{\bf q}\,^{\prime}\,)\circledast_\varepsilon{\bf q}\,^{\prime\prime}-{\bf q}\circledast_\varepsilon({\bf q}\,^{\prime}\circledast_\varepsilon{\bf q}\,^{\prime\prime}\,) = 0 \ .
\end{equation}
\item[(V4) ] ${\bf q}\circledast_\varepsilon{\bf 0}={\bf 0}\circledast_\varepsilon{\bf q}={\bf q}$, and $(-{\bf q})\circledast_\varepsilon(-{\bf q}\,^{\prime})=-{\bf q}\circledast_\varepsilon{\bf q}\,^{\prime}$.
\end{description}
The above properties are essential for the definition of the $su(2)$-like star product for the functions on $\mathbb{R}^3$, see \cite{OR13,KVit}. 

The standard three-dimensional cross product $\times_\varepsilon$ is related to the algebra of quaternions $\mathbb{H}$ in the same way as the seven-dimensional cross product $\times_\eta$ to the algebra of octonions $\mathbb{O}$. For vectors $\vec
k=(k^A)$ and $\vec k'=(k^{\prime\,A})$ from $\mathbb{R}^7$ it is defined by,
\begin{equation}
(\vec k\,\times_\eta\,\vec k'\,)^A := \eta^{ABC}\, k^B\, k^{\prime\,C}\,.
\label{eq:7dvectorprod}
\end{equation}
 For better understanding of the underlying
geometry of the seven-dimensional cross
product, it is useful to note that it can be expressed in terms of the
algebra of imaginary octonions by writing $X_{\vec k}:=k^A\, e_A$ and observing that
\begin{equation}
\label{eq:Xveckcomm}
X_{\vec k\,\times_\eta\,\vec k'}=\mbox{$\frac12$}\, \big[X_{\vec k},X_{\vec k'} \big] \ .
\end{equation}
Consider the splitting of the
seven-dimensional vector $\vec k$ according to the components $(e_i,f_i,e_7)$ of the previous
Section, $\vec k=( {\bf l},  {\bf k},k_4)$, where
$ {\bf l}=(l^i), {\bf k}=(k_i)\in\mathbb{R}^3$. With respect to this decomposition, by using (\ref{eq:Xveckcomm}) and (\ref{oct2a}) the seven-dimensional cross product can be written in terms of the three-dimensional cross product as
\begin{eqnarray}\label{oct12}
\vec k\,\times_\eta\,\vec k'=\big( {\bf l}\,\times_\varepsilon\, {\bf l'}- {\bf k}\,\times_\varepsilon\, {\bf k'} +
  k_4'\,  {\bf k}-k_4\, {\bf k'} \,,\,  {\bf k}\,\times_\varepsilon\, {\bf l'}- {\bf l}\,\times_\varepsilon\,
  {\bf k'}+k_4\,  {\bf l'}-k_4'\,  {\bf l} \,,\,  {\bf l}\,\cdot\, {\bf k'}- {\bf k}\,\cdot\,
   {\bf l'}\,\big) \ .
\end{eqnarray}
In particular, reduction to the three-dimensional subspace spanned by
$e_i$ gives
\begin{equation}\label{oct12b}
( {\bf l}, {\bf 0},0)\,\times_\eta\,( {\bf l'}, {\bf 0},0) = ( {\bf l}\,\times_\varepsilon\, 
{\bf l'}, {\bf 0},0) \ ,
\end{equation}
yielding the expected three-dimensional cross product. 

The defining properties of cross products are\begin{description}
\item[(C1)]  $\vec k\,\times_\eta\, \vec k'=-\vec k'\,\times_\eta\, \vec k$\, ;
\item[(C2)] $\vec k\,\cdot\,(\vec k'\,\times_\eta\, \vec k^{\prime\prime}\,)=-\vec k'\,\cdot\,(\vec k\,\times_\eta\, \vec k^{\prime\prime}\,)$\, ;
\item[(C3)]  $|\vec k\,\times_\eta\,\vec k'\,|^2 = |\vec k|^2\, |\vec
  k' \,|^2 - (\vec k\,\cdot\, \vec k'\,)^2$, where $|\vec k|=\sqrt{\vec k\,\cdot\,\vec k}$ is the Euclidean vector norm.
 \end{description}
As usual property {\bf{(C1)}} is equivalent to the statement that the
cross product $\vec k\,\times_\eta\, \vec k'$ is non-zero if and only if
$\vec k$, $\vec k'$ are linearly independent vectors, property
{\bf{(C2)}} is equivalent to the statement that it is orthogonal to
both $\vec k$ and $\vec k'$, while property {\bf{(C3)}} states that
its norm calculates the area of the triangle spanned by $\vec k$
and $\vec k'$ in $\mathbb{R}^7$. However, unlike the three-dimensional cross product $\,\times_\varepsilon\,$, due to \eqref{oct3} it does not obey the Jacobi identity: Using \eqref{epsilon7} the Jacobiator is given by
\begin{eqnarray}
\vec{J}_\eta(\vec k,\vec k ',\vec k^{\prime\prime}\,)&:=& (\vec k \,\times_\eta\, \vec k'\,)\,\times_\eta\,\vec k^{\prime\prime}+(\vec k' \,\times_\eta\, \vec k^{\prime\prime}\,)\,\times_\eta\,\vec k+(\vec k \,\times_\eta\, \vec k^{\prime\prime}\,)\,\times_\eta\,\vec k' \nonumber\\[4pt]
&=& 3\,\big(\,(\vec k \,\times_\eta\, \vec k'\,)\,\times_\eta\,\vec k^{\prime\prime}+(\vec k'\,\cdot\,\vec k^{\prime\prime}\,)\ \vec k -(\vec k\,\cdot\,\vec k^{\prime\prime}\,)\ \vec k'\, \big) \ ,
\label{eq:Jaceta}
\end{eqnarray}
which can be represented through the associator on the octonion algebra
$\mathbb{O}$ as
\begin{equation}
\nonumber
X_{\vec{J}_\eta(\vec k,\vec k ',\vec k^{\prime\prime}\,)} =
\mbox{$\frac14$}\, \big[X_{\vec k},X_{\vec k'},X_{\vec k''}\big] =
\mbox{$\frac32$}\, \big((X_{\vec k}\,X_{\vec k'})\,X_{\vec k''}-
X_{\vec k}\,(X_{\vec k'}\,X_{\vec k''} ) \big) \ .
\end{equation}
From (\ref{eq:Jaceta}) one may easily see that the Jacobiator ${\bf J}_\varepsilon$ on the corresponding three-dimensional subspace now vanishes, providing the well known identity for the cross product in three dimensions.

Using the properties of the seven-dimensional cross product we may generalize the three-dimensional vector star sum (\ref{ds3}) to the seven-dimensional case corresponding to the algebra of imaginary octonions. For any pair of vectors $\vec p,\vec p\,^{\prime}$ from the unit ball $B^7\in\mathbb{R}^7$, $|\vec p\,|\leq1$, define
\begin{equation}\label{vstar}
\vec p\circledast_\eta\vec p\,^{\prime}=\ \epsilon_{\vec p,\vec
  p\,'}\,\big(\, \sqrt{1-| \vec p\,^{\prime}|^2}\,\,\vec p+ {\sqrt{1-|
    \vec p\,|^2} }\,\,\vec p\,^{\prime}-\vec p\,\times_\eta\, \vec
p\,^{\prime}\, \big) \ ,
\end{equation}
where $\epsilon_{\vec p_1,\vec p_2}=\pm\,1$ is the sign of $\sqrt{1-|\vec p_1|^2}\, \sqrt{1-|\vec p_2|^2} -\vec p_1\,\cdot\, \vec p_2$. 
Using {\bf{(C1)}}--{\bf{(C3)}} we find
    \begin{equation}\label{eq:vstarid}
1-|\vec p\circledast_\eta \vec p\,^{\prime}\,|^2=\big(\, \sqrt{1-|\vec  p\, |^2}\, \sqrt{1-|\vec p\,^{\prime}\, |^2}-\vec p\,\cdot\,\vec p\,^{\prime}\, \big)^2\geq0 \ ,
\end{equation}
meaning that the vector $\vec p\circledast_\eta\vec p\,^{\prime}$ also belongs to
the unit ball $B^7$. Just like in the three-dimensional case, it admits an identity element given by the zero vector in $B^7$,
$
\vec p \circledast_\eta\vec 0= \vec p = \vec 0 \circledast_\eta \vec p
\ ,
$
and the inverse of $\vec p\in B^7$ is $-\vec p\in B^7$, i.e.,
$
\vec p \circledast_\eta(-\vec p\,)=\vec 0= (-\vec p\,)\circledast_\eta
\vec p \ .
$
The commutator reproduces the seven-dimensional cross product,
\begin{equation}\label{eq:vstarcomm}
\vec p\circledast_\eta\vec p\,^{\prime}- \vec p\,^{\prime}
\circledast_\eta\vec p = -2\, \vec p \,\times_\eta\, \vec
p\,^{\prime} \ .
\end{equation}
Three-dimensional vector star product was associative (\ref{vssass}), what is not the case of the seven-dimensional on. However, taking into account the definition of the sign factor $\epsilon_{\vec p_1,\vec p_2}$ one may check that the corresponding associator
is related to the Jacobiator \eqref{eq:Jaceta} for the
cross product \eqref{eq:7dvectorprod} through
\begin{eqnarray}\label{assvstar}
\vec A_\eta(\vec p,\vec p\,^{\prime},\vec p\,^{\prime\prime}\,):= (\vec p\circledast_\eta\vec p\,^{\prime}\,)\circledast_\eta\vec p\,^{\prime\prime}-\vec p\circledast_\eta(\vec p\,^{\prime}\circledast_\eta\vec p\,^{\prime\prime}\,) = \mbox{$\frac23$}\, \vec J_\eta(\vec p,\vec p\,^{\prime},\vec p\,^{\prime\prime}\,) \ .
\end{eqnarray}
The components of the
associator (\ref{assvstar}) can be represented as
\begin{equation*}
A_\eta(\vec p,\vec p\,^{\prime},\vec p\,^{\prime\prime}\, )^A= \eta^{ABCD}\, p^B\, p^{\prime\, C}\, p^{\prime\prime\, D} \ .
\end{equation*}
It is non-vanishing but totally antisymmetric, which means that the
seven-dimensional vector star sum (\ref{vstar}) is nonassociative but
alternative. The reduction of the seven-dimensional vector star sum
(\ref{vstar}) on this three-dimensional subspace reproduces the
three-dimensional vector star sum $\circledast_\varepsilon$,
\begin{equation}\label{vstar3}
({\bf q},{\bf 0},0)\circledast_\eta ({\bf q}',{\bf 0},0)=({\bf
  q}\circledast_\varepsilon {\bf q}',{\bf 0},0) \ ,
\end{equation}
which by \eqref{assvstar} and (\ref{eq:Jaceta}) is now associative.

As well as in the case of the seven-dimensional cross product the underlying geometry of the vector star sum can be better understood providing  an
alternative eight-dimensional characterisation of the binary operation
\eqref{vstar}. For this, we
note that for arbitrary vectors $P,P'\in \mathbb{R}^8$ one has~\cite{Salamon2010}
\begin{eqnarray}
\label{eq:XP8D}
X_P\, X_{P'} = \big(p_0\,p_0'-\vec p \,\cdot\, \vec p\,'\,\big)\,{\bf1}
+p_0\, X_{\vec p\,'}+p_0'\, X_{\vec p} +X_{\vec p\,\times_\eta\,
  \vec p\,'} \ .
\end{eqnarray}
By restricting to vectors $P$ from the unit sphere $S^7\subset \mathbb{R}^8$,
i.e., $|P|=1$, we
can easily translate this identity to the vector star sum
\eqref{vstar}: We fix a hemisphere $p_0=\pm\, \sqrt{1-|\vec p\,|^2}$,
such that for vectors $\vec p,\vec p\,'$ in the ball $B^7 $ we
reproduce the seven-dimensional vector star sum through
\begin{eqnarray}
\label{1}
X_{\vec p\circledast_\eta\vec
  p\,^{\prime}}=\mathfrak{Im}\big( X_{P\,^{\prime}}\, X_P\big) =
\mbox{$\frac12$}\, \big(X_{P\,^{\prime}}\, X_{P}-\bar X_{P}\, \bar X_{P\,^{\prime}}\big) \ ,
\end{eqnarray}
where the sign factor $\epsilon_{\vec p,\vec p\,'}$ from
\eqref{vstar}, which is the sign of the real part of \eqref{eq:XP8D}, ensures that the result of the vector star sum remains in
the same hemisphere. This interpretation of the vector star sum is
useful for deriving various properties. For example, since the algebra of
octonions $\mathbb{O}$ is a normed algebra, for $|P|=1$ we have $|X_P|=1$
and
\begin{eqnarray}
\nonumber
\big|X_{P\,^{\prime}}\,
X_{P}\big|^2=\big|\mathfrak{Re}(X_{P\,^{\prime}}\, X_{P})
\big|^2+\big|\mathfrak{Im}(X_{P\,^{\prime}}\, X_{P}) \big|^2=1 \ .
\end{eqnarray}
From \eqref{eq:XP8D} and \eqref{1} we then immediately infer the
identity \eqref{eq:vstarid}.

To extend the structure (\ref{vstar}) over the entire vector space $V$ we introduce the map
\begin{equation}\label{eq:pkmap}
 \vec p=\frac{\sin(\hbar\, |\vec k|)}{|\vec k|}\ \vec k \qquad \mbox{with} \quad k^A\in\mathbb{R} \ .
\end{equation}
The inverse map is given by
\begin{equation}\nonumber
 \vec k=\frac{\sin^{-1}|\vec p\,|}{\hbar\, |\vec p\,|}\ \vec p \ .
\end{equation}
Then for each pair of vectors $\vec k,\vec k'\in V$, we define the \emph{deformed vector sum}:
\begin{equation}
\label{Bk}
\vec{\mathcal{ B}}_\eta(\vec k,\vec k'\,):= \mbox{$\frac1\hbar$}\, \vec{\mathcal{ B}}^{\,\prime}_\eta(\hbar\, \vec k,\hbar\, \vec k'\,)= \left.\frac{\sin^{-1}|\vec p\circledast_\eta\vec p\,'\,|}{\hbar\,|\vec p\circledast_\eta\vec p\,'\,|}\ \vec p\circledast_\eta\vec p\,'\, \right|_{ \vec p=\vec k\sin(\hbar\, |\vec k|)/|\vec k|} \ .
\end{equation}
From the definition (\ref{Bk}) and the properties of the operation $\circledast_\eta$ one finds immediately that:
\begin{description}
\item[(B1)]  $\vec{\mathcal{ B}}_\eta(\vec k,\vec k'\,)=-\vec{\mathcal{ B}}_\eta(-\vec k',-\vec k\,)$\, ;
\item[(B2)]  $\vec{\mathcal{ B}}_\eta(\vec k,\vec 0\,) =\vec k=\vec{\mathcal{ B}}_\eta(\vec 0,\vec k\,)$\, ;
\item[(B3)]  Perturbative expansion: \ $\vec{\mathcal{ B}}_\eta(\vec k,\vec k'\,)=\vec k+\vec k'-{2\, \hbar}\, \vec k\,\times_\eta\,\vec k'+O(\hbar^2)$\, ;
\item[(B4)] The associator $$\vec{\mathcal{A}}_\eta(\vec k,\vec k',\vec k^{\prime\prime}\,):=\vec{\mathcal{ B}}_\eta\big(\vec{\mathcal{ B}}_\eta(\vec k,\vec k'\,)\,,\,\vec k^{\prime\prime}\,\big)-\vec{\mathcal{ B}}_\eta\big(\vec k\,,\,\vec{\mathcal{ B}}_\eta(\vec k',\vec k^{\prime\prime}\,)\big)$$ is antisymmetric in all arguments. \end{description}

\subsection{Octonionic star product}

Let us define the star product as
\begin{equation}\label{w1}
(f\star_\eta g)( \vec \xi\ ) =\int \, \frac{d^{7}\vec k}{( 2\pi
) ^{7}} \ \frac{d^{7}\vec k'}{( 2\pi
) ^{7}} \ \tilde{f}( \vec k\,)\, \tilde{g}(\vec k'\,)\, e^{i\vec{\mathcal{ B}}_\eta(\vec k,\vec k'\,)\,\cdot\,{\vec \xi}} \ ,
\end{equation}
where again $\tilde{f}$ stands for the Fourier transform of the
function $f$ and $\vec{\mathcal{ B}}_\eta(\vec k,\vec k'\,)$ is the deformed vector sum \eqref{Bk}. By definition it is the Weyl star product.

Due to the properties {\bf{(B1)}} and {\bf{(B2)}} of the deformed vector addition $\vec{\mathcal{ B}}_\eta(\vec k,\vec k'\,)$, this star product is Hermitean, $(f\star_\eta g)^\ast=g^\ast \star_\eta f^\ast$,
and unital, $f\star_\eta 1=f=1\star_\eta f$. It can be regarded as
a quantization of the dual of the pre-Lie algebra \eqref{oct2}
underlying the octonion algebra $\mathbb{O}$; in particular, by property
{\bf{(B3)}} it provides a quantization of the quasi-Poisson bracket (\ref{oct2a}): Defining $[f,g]_{\star_\eta}=f\star_\eta g-g \star_\eta f$, we have
\begin{equation}\label{41}
       \lim_{\hbar\to0}\, \frac{[f,g]_{\star_\eta}}{i\hbar}=2\, \xi_A\, \eta_{ABC}\, \partial^Bf\, \partial^Cg = \{f,g\}_\eta \ .
\end{equation}
Property {\bf{(B4)}} implies that the star product (\ref{w1}) is alternative on monomials and plane waves.

Let us calculate $\xi_A\star_\eta f$ explicitly using (\ref{w1}). We have
\begin{eqnarray}
 \xi_A\star_\eta f =-\int\, \frac{d^{7}\vec k'}{( 2\pi
) ^{7}} \ \xi_D\, \frac{ \partial  \mathcal{ B}_\eta(\vec k,\vec k'\, )^D}{\partial k^A}\bigg|_{\vec k=\vec 0}\ \tilde{f}(\vec k'\,)\, e^{i\vec{\mathcal{ B}}_\eta(\vec 0,\vec k'\, )\,\cdot\,{\vec  \xi}}\nonumber
\end{eqnarray}
and after some algebra one finds
\begin{eqnarray*}
\frac{ \partial \mathcal{ B}_\eta(\vec k,\vec k'\, )^D}{\partial k^A}\bigg|_{\vec k=\vec 0}= -\hbar \, \eta_{ADE}\, k^{\prime\, E}+\delta_{AD}\, \hbar\, |\vec k'\,|\cot(\hbar\, |\vec k'\,|)
+\frac{k^{\prime}_{A}\, k'_D}{|\vec k'\, |^2}\, \Big(\hbar\, |\vec k'\, |\cot(\hbar\, |\vec k'\,|)-1\Big) \ .\nonumber
\end{eqnarray*}
Taking into account property {\bf{(B2)}} and integrating over $\vec k'$, we arrive at
\begin{eqnarray}\label{poly}
 \xi_A\star_{\eta} f &=& \Big(\xi_A+i\hbar\, \eta_{ABC} \,
                         \xi_C\, \partial^B \\ && \qquad +\, 
  \big( \xi_A\, {\triangle_{\vec\xi}}- (\vec\xi\,\cdot\,
                                                  \nabla_{\vec\xi}\, ) \, \partial_A
                                                  \big)\,
                                                  \triangle_{\vec\xi}^{-1}\,
                                                  \big( \hbar\,
                                                  \triangle_{\vec\xi}^{1/2}
                                                  \, \coth(\hbar\,
                                                  \triangle_{\vec\xi}^{1/2}
                                                  \,)-1\big)\Big)\triangleright
                                                  f(\vec\xi\ ) \nonumber
\end{eqnarray}
where ${\triangle_{\vec\xi}}=\nabla_{\vec\xi}^2= \partial_A\, \partial^A$ is the flat space Laplacian in seven dimensions. In particular for the Jacobiator one finds
\begin{equation}\label{oct5}
[ \xi_A, \xi_B, \xi_C]_{\star_\eta} =12\, \hbar^2\, \eta_{ABCD} \, \xi_D \ ,
\end{equation}
which thereby provides a quantization of the classical 3-brackets
\eqref{mp2}.

Setting $e_A=0$ for $A=4,5,6,7$ (equivalently $f_i=e_7=0$) reduces the
nonassociative algebra of octonions $\mathbb{O}$ to the associative algebra of quaternions $\mathbb{H}$,
whose imaginary units $e_i$ generate the $su(2)$ Lie algebra
$[e_i,e_j]=2\,\varepsilon_{ijk}\, e_k$. The corresponding reduction of the deformed vector sum \eqref{Bk} reproduces the
three-dimensional vector sum ${\bf{\mathcal{ B}}}_\varepsilon({\bf l},{\bf l}'\, )$ from~\cite{OR13,KVit},
\begin{equation}
\label{Bk3}
\vec{\mathcal{ B}}_\eta\big( ({\bf l},{\bf 0},0)\,,\,({\bf l}',{\bf
  0},0) \big)=\big({\bf{\mathcal{ B}}}_\varepsilon({\bf l},{\bf l}'\, ),{\bf
  0},0 \big) \ ,
\end{equation}
with vanishing associator {\bf{(B4)}}. From (\ref{Bk3}) it follows that, for functions $f,g$ on this three-dimensional
subspace, the
corresponding star product $(f\star_\varepsilon g)(\bf\xi,\bf0,0)$
from (\ref{w1}) reproduces the associative star product of~\cite{OR13,KVit} for the quantization
of the dual of the Lie algebra $su(2)$. 

Note that neither the octonionic star product $\star_\eta$ defined in (\ref{w1}), nor the $su(2)$-like one $\star_\varepsilon $ are closed with respect to the integral, i.e. do not satisfy (\ref{cust1}). The solution to this problem was proposed in \cite{KVit}, by an appropriate choice of a gauge transformation \begin{equation}\label{i7}
f\bullet_\eta g= \mathcal{D}^{-1}\big( \mathcal{D}f\star_\eta \mathcal{D}g \big) \qquad \mbox{with} \quad \mathcal{D}=1+{O}(\hbar) \ ,
\end{equation}  one may construct a closed star product satisfying
 \begin{equation}\label{i1a}
 \int \, d^{7}\vec x\ f\bullet_\eta g = \int\, d^7\vec x\ f\, g \ .\end{equation}
 For the octonionc star product (\ref{w1}) the corresponding gauge transformation was found in \cite{Kup24} and has a form:
\begin{equation}\label{oct13}
\mathcal{D}=\Big( \big(\hbar\,{\bf\triangle}_{\vec \xi}^{1/2}\big)^{-1} \sinh\big(\hbar\, {\bf\triangle}_{\vec \xi}^{1/2}\big) \Big)^{6} \ .
\end{equation}
As it was already mentioned in the introduction, the gauge equivalent star product (\ref{i7}) with (\ref{oct13}) provides the quantization of the same algebra (\ref{oct2a}) as the original one (\ref{w1}). One may also show, see \cite{KS}, that the closed star product $\bullet_\eta$ on the Schwartz functions satisfies the 3-cyclicity property (\ref{aust1}).

\subsection{Phase space star product for M-theory $R$-flux background}

Using the
closed octonionic star product $\bullet_\eta$ obtained in the previous subsection we will now construct the quantization of the quasi-Poisson structure defined in \eqref{RM}. For this, we define a star product of functions
on the seven-dimensional M-theory phase space by the prescription
\begin{equation}
\label{oct10}
(f\bullet_\lambda g)(\vec x\, )=(f_{\mit\Lambda}\bullet_\eta g_{\mit\Lambda})( \vec\xi\ )\big|_{\vec\xi={\mit\Lambda}^{-1}\,\vec x}\,,
\end{equation}
where $f_{\mit\Lambda}(\vec\xi\ ):= f({\mit\Lambda}\,\vec\xi\ )$. Using the deformed vector sum $\vec{\mathcal{ B}}_\eta(\vec k, \vec k'\,)$ from (\ref{Bk}), and the gauge transformation (\ref{oct13}) we can write \eqref{oct10} as
\begin{eqnarray}\label{i14}
(f\bullet_\lambda g)(\vec x\, ) &=& \int\, \frac{d^{7}\vec k}{( 2\pi
) ^{7}} \ \frac{d^{7}\vec k'}{( 2\pi
) ^{7}} \ \tilde{f}( \vec k\, )\, \tilde{g}( \vec k'\, )\, e^{i\vec{\mathcal{ B}}_\eta({\mit\Lambda}\,\vec k,{\mit\Lambda}\, \vec k'\, )\,\cdot\, {\mit\Lambda}^{-1}\,\vec x} \\ && \qquad\qquad\qquad\qquad \times \ \Big(\, \frac{\sin\big(\hbar\, |{\mit\Lambda}\,\vec k|\big)\sin\big(\hbar\, |{\mit\Lambda}\,\vec k'\, |\big)}{\hbar\, |{\mit\Lambda}\,\vec k| \, |{\mit\Lambda}\,\vec k'\, |} \, \frac{|\vec{\mathcal{ B}}_\eta({\mit\Lambda}\,\vec k,{\mit\Lambda}\, \vec k'\, )|}{\sin\big(\hbar\, |\vec{\mathcal{ B}}_\eta({\mit\Lambda}\,\vec k,{\mit\Lambda}\, \vec k'\, )|\big)}\, \Big)^6 \ .\notag
\end{eqnarray}
This star product provides a quantization of the brackets (\ref{oct4a}), since
\begin{equation}\label{i15}
\lim_{\stackrel{\scriptstyle\hbar,\ell_s\to0}{\scriptstyle \ell_s^3/\hbar^2={\rm constant}}} \, \frac{[f,g]_{\bullet_\lambda}}{i\hbar} =\{f,g\}_{\lambda} \ .
\nonumber\end{equation}

To study the contraction limit $\lambda\to0$ one needs to calculate,
\begin{eqnarray}
\nonumber
\lim_{\lambda\to0} \, \vec{\mathcal{ B}}_\eta({\mit\Lambda}\,\vec
k,{\mit\Lambda}\, \vec k'\, )\,\cdot\, {\mit\Lambda}^{-1}\,\vec x &=& ( k+ k'\,)
\,\cdot\, x + (k_4+k_4'\,)
\, x^4 + ( l+ l'\,)
\,\cdot\, p \\
&& -\, \mbox{$\frac1{2\hbar}$}\,\big(\ell_s^3\, R\,  p\,\cdot\, (
k\,\times_\varepsilon\, k'\,)+\hbar^2\, x^4\, ( k\,\cdot\, l'- l\,\cdot\, k'\,)
\big) \ .
\label{l10}
\end{eqnarray}
Up to the occurance of the circle fibre coordinate $x^4$, this
expression coincides exactly with (\ref{BR}). In the dimensional reduction of M-theory to IIA string theory
we restrict the algebra of functions to those which are constant along
the $x^4$-direction; they reduce the Fourier space integrations in
\eqref{i14} to the six-dimensional hyperplanes $k_4=k_4'=0$. The coordinate $x^4$ is central in the star product algebra of
functions in the limit $\lambda\to0$; as before we may therefore set
it to any non-zero constant value, which we take to be
$x^4=1$.
The extra factors in \eqref{i14} are simply unity in the contraction limit $\lambda\to0$, and so the star product \eqref{i14} dimensionally reduces to \eqref{starrb1},
\begin{equation}
\lim_{\lambda\to0} \, (f\bullet_\lambda g)(\vec x\, ) = (f\star g)(x) \ .
\end{equation}
It is Hermitean, $(f\bullet_\lambda g)^\ast=g^\ast \bullet_\lambda
f^\ast,$ unital, $f\bullet_\lambda 1=f=1\bullet_\lambda f$, closed with respect to the integral and 3-cyclic, \begin{equation}
    \int\, d^7\vec x\ (f\bullet_\lambda g)\bullet_\lambda h= \int \, d^7\vec x\ f\bullet_\lambda (g\bullet_\lambda h) \ .
\label{i21}\end{equation}
This property can be regarded as the absence on-shell of
nonassociativity (but not noncommutativity) among cubic interactions
of fields.

\section{Minimal set of requirements}

Here we discuss the reasonable physical conditions that one may impose on the non-associative star products. The both star products discussed in this notes are Hermitean and unital. The Hermiticity reflects the reality of physical observables, while the unitality is essential for the definition of algebra. Taking the limit $R\to0$ in the star product (\ref{starrb}) one recovers the standard associative Moyal product. In case of the octonionic star product (\ref{w1}), in section 4.2 it was shown that restricting the functions $f,g$ on the three-dimensional
subspace endowed with the $su(2)$-structure, one obtains the associative star product $(f\star_\varepsilon g)(\bf\xi,\bf0,0)$, of~\cite{OR13,KVit} for the quantization
of the dual of the Lie algebra $su(2)$. These observations are in agreement with the topological limit in open string theory \cite{CoSch,MK1,Herbst}, basically meaning that the star product should become associative if the corresponding bi-vector $\theta$ is a Poisson one. 

In \cite{KV15} it was proposed an iterative procedure that allows one to compute unital, Heritean, Weyl (\ref{weyl}) star product for any given quasi-Poisson structure $\theta^{ij}(x)$ up to any desired order in $\hbar$. Up to the third order it reads,
\begin{eqnarray}
&&(f\star g)(x)=f\cdot g+i\hbar \theta ^{ij}\partial
_{i}f\partial _{j}g  \notag \\
&&-\frac{\hbar ^{2}}{2}\theta^{ij}\theta ^{kl}\partial _{i}\partial
_{k}f\partial _{j}\partial _{l}g-\frac{\hbar ^{2}}{3}\theta ^{ij}\partial
_{j}\theta ^{kl}\left( \partial _{i}\partial _{k}f\partial _{l}g-\partial
_{k}f\partial _{i}\partial _{l}g\right) \notag\\
&&-i\hbar^3\left[\frac{1}{3}\theta ^{nl}\partial _{l}\theta ^{mk}\partial _{n}\partial
_{m}\theta ^{ij}\left( \partial _{i}f\partial _{j}\partial _{k}g-\partial
_{i}g\partial _{j}\partial _{k}f\right)\right.   \notag \\
&&+ \frac{1}{6}  \theta^{nk}\partial_n\theta^{jm}\partial_m\theta^{il}
\left(\partial_i\partial_jf\partial_k\partial_lg-\partial_i\partial_jg\partial_k\partial_lf\right)   \notag \\
&&+\frac{1}{3}\theta ^{ln}\partial _{l}\theta ^{jm}\theta ^{ik}\left(
\partial _{i}\partial _{j}f\partial _{k}\partial _{n}\partial _{m}g-\partial
_{i}\partial _{j}g\partial _{k}\partial _{n}\partial _{m}f\right)
\notag \\
&&+\frac{1}{6}\theta ^{jl}\theta ^{im}\theta ^{kn}\partial _{i}\partial
_{j}\partial _{k}f\partial _{l}\partial _{n}\partial _{m}g  \notag \\
&& +\left.\frac{1}{6}\theta ^{nk}\theta ^{ml}\partial _{n}\partial _{m}\theta
^{ij}\left( \partial _{i}f\partial _{j}\partial _{k}\partial _{l}g-\partial
_{i}g\partial _{j}\partial _{k}\partial _{l}f\right)\right]+\mathcal{O}\left( \hbar ^{4}\right)~.   \label{star3}
\end{eqnarray}
To this order the star product (\ref{star3}) is compatible with the topological limit, since the corresponding associator,
\begin{eqnarray}
&& A(f,g,h)={2\hbar^2}\ \Pi^{ijk} \partial_i f \partial_j g \partial_k h\nonumber\\
&&-\frac{i\hbar^3}{24}\left[\left( \theta^{mi}\partial_m\Pi^{jkl}+\frac12\ \Pi^{ikm}\partial_m\theta^{lj}-\frac12\ \Pi^{ilm}\partial_m\theta^{kj}\right)\, \left(\partial_i\partial_j f\partial_kg\partial_l  h+\partial_i\partial_j h\partial_kg\partial_l  f\right)\right.\nonumber\\
&&+\left(\Pi^{lkm}\partial_m\theta^{ij}+\Pi^{ikm}\partial_m\theta^{lj}\right)\ \partial_if\partial_j\partial_kg\partial_lh\nonumber\\
&&-\left.2\theta^{ik}\Pi^{jlm}\, \left(\partial_i\partial_jf\partial_k\partial_lg\partial_mh-\partial_i\partial_jf\partial_mg\partial_k\partial_lh+\partial_i\partial_jh\partial_k\partial_lg\partial_mf\right)\right]+\mathcal{O}\left( \hbar ^{4}\right)~.\label{A3}
\end{eqnarray}
is proportional to $ \Pi^{ijk}$, and consequently vanishes if $\theta$ is Poisson. To make an iterative procedure \cite{KV15} compatible with the topological limit in higher orders one needs to introduce ``quantum'', i.e., proportional to the powers of $\hbar$, corrections to the original bi-vector $\theta^{ij}$, corresponding to the Kontsevich diagrams with wheels \cite{Dito}. The calculation of such a corrections is a highly non-trivial task \cite{Penkava,KV08}. The first non-vanishing contribution appears in the order $\hbar^2$ and the resulting bi-vector reads,
 \begin{equation}
\theta_{quant}^{jk}=\theta^{jk}+\frac{\hbar^2}{6}\left(\partial_m\theta^{nl}\partial_n\theta^{mi}\partial_l\partial_i\theta^{jk}-
    2\theta^{il}\partial_n\partial_i\theta^{jm}\partial_m\partial_l\theta^{kn}\right)+\mathcal{O}\left( \hbar ^{4}\right)\,. \label{qc}
\end{equation}

Two other important properties which the both star products from the Sections 2 and 4 enjoy are the conditions of closure with respect to the appropriate integration measure (\ref{cust1}) and the 3-cyclicity (\ref{aust1}). The latter means that the associativity holds true up to the surface terms. These properties are essential for the formulation of the non-associative quantum mechanics \cite{MSS2,Szabo:2017yxd}.  In string theory the property of 3-cyclicity can be interpreted as the requirement that the non-associativity should not manifest itself in on-shell scattering amplitudes, but only off-shell, see \cite{Blum3} for more details.

The star product (\ref{star3}) just like the octonionic star product (\ref{w1}) does not satisfy the closure condition (\ref{cust1}). However, one may follow the same logic as in the section 4.2 and search for the gauge equivalent closed star product. At least in the associative case the existence of the closed star product for arbitrary Poisson $\theta$ was proven in \cite{Felder:2000nc}. For non-associative star product (\ref{star3}) the leading order of the corresponding gauge transformation was constructed in \cite{Kup24}. 

The condition of 3-cyclicity (\ref{aust1}) for arbitrary $\theta$ seems to be less trivial. In \cite{Herbst} it was shown that the closed non-associative star product compatible with the topological limit is 3-cyclic up to the second order in the derivative expansion of the non-commutativity parameter $\theta$. Nevertheless this does not prove yet the existence of the 3-cyclic star products for any $\theta$ in all orders. Since, the expression for the associator (\ref{A3}) already in the third order in the deformation parameter $\hbar$ contains the terms with two derivatives on $\theta$, like $\theta\partial\Pi$ or $\Pi\partial\theta$, the integration by parts over $f$, $g$ or $h$ may lead to the non-vanishing terms with three or more derivatives on $\theta$, which were not taken into account in \cite{Herbst}. We believe that more careful analysis of the 3-cyclicity condition is needed. For some quasi-Poisson structures $\theta$ it definitely holds true, but possibly not for all.

To conclude let us formulate the minimal set of the physically motivated requirements which can be used instead of the associativity condition (\ref{ass}) to restrict the higher order terms of the formal expression (\ref{star}) for the  star product representing quantization of a given quasi-Poisson structure $\theta^{ij}(x)$:
\begin{description}
\item[(a$\star$)] Hermiticity:
\begin{equation}
(f\star g)^*=g^*\star f^* \ .
\end{equation}
\item[(b$\star$)] Unitality:
\begin{equation}
1\star f=f=f\star 1 \ .
\end{equation}
\item[(c$\star$)] Topological limit.
\end{description}

\subsection*{Acknowledgments}

We thank Dieter L\"ust, Patrizia Vitale, Peter Schupp, Ralph Blumenhagen and  Richard Szabo for helpful discussions.
Also we would like to thank the organizers of the {\it Workshop on  Noncommutative Field Theory and Gravity} at the
Corfu Summer Institute 2017, for the nice and productive atmosphere. The author acknowledges the CAPES-Humboldt Fellowship No.~0079/16-2 and CNPq Grant No.~305372/2016-5.

\end{document}